\documentclass[a4paper]{article}

\usepackage{icrc2013}

\title{Diffusion coefficient and radial gradient of galactic cosmic rays}

\shorttitle{Diffusion coefficient and radial gradient of galactic cosmic rays}

\authors{
Renata Modzelewska
}

\afiliations{
Institute of  Math. and Physics, Siedlce University, Siedlce, Poland
}

\email{renatam@uph.edu.pl}

\abstract{We present the temporal changes of the diffusion coefficient $K$ of galactic cosmic rays (GCRs) at the Earth orbit calculated based on the experimental data using two different methods. The first approach is based on the Parker convection-diffusion approximation of GCR modulation \cite{bib:Parker65}: i.e. $K \sim Vr/dI$  where $dI$  is the variation of the GCR intensity measured by neutron monitors (NM), $V$ is the solar wind velocity  and  $r$ is the radial distance. The second approach is based on the interplanetary magnetic field (IMF) data. It was suggested that parallel mean free path $\lambda_{\parallel}$  can be expressed in terms of B as follows \cite{bib:zank98}-\cite{bib:bazilevskaya13}: $\lambda_{\parallel}\propto \frac{B^{\frac{5}{3}}}{\delta B^{2}}$, where $\delta B$  is the standard deviation. In our calculations we used an approximately equivalent expression   $\lambda_{\parallel}\propto \frac{B}{\delta B}$.  Using data of the product $\lambda_{\parallel} \nabla_{r} n$ of the parallel mean free path  $\lambda_{\parallel}$ and radial gradient $\nabla_{r} n$  of GCR calculated based on the GCR  anisotropy data (Ahluwalia et al., this conference ICRC 2013, poster ID: 487 \cite{bib:Ahl13}), we estimate the temporal changes of the radial gradient $\nabla_{r} n$  of GCR at the Earth orbit. We show that the radial gradient $\nabla_{r} n$  exhibits a strong solar cycle dependence (11-year variation) and a weak solar magnetic cycle dependence (22-year variation), being in agreement with the previous other calculations and with PIONEER/VOYAGER observations.}


\begin{document}
\maketitle

\section{Introduction}

The galactic cosmic rays (GCRs) transport in the heliosphere is governed by the four important processes: outward convection by the solar wind, inward diffusion, particle drifts (gradient, curvature and on the neutral sheet) in the turbulent interplanetary magnetic field (IMF) and  adiabatic cooling. Estimation of the local electromagnetic conditions near the Earth orbit is possible by establishing modulation parameters as diffusion coefficient, density gradients etc. It is especially essential when in situ measurements are absent. It is the basic knowledge needed to study space weather prediction. A major issue in GCR transport research and space weather studies is how GCR particles propagate through the heliosphere, and how interact with the interplanetary space especially in the inner heliosphere near the Earth's orbit. The essential significance of characterizing GCR propagation is evident, because this will lead to a practical capability in space weather forecasting which has important consequences for life and technology on the Earth and also in the interplanetary space.

The first theoretical description of cosmic ray transport coefficients was done by Jokipii  \cite{bib:jokipii66} by formulation the quasilinear theory (QLT) for GCR diffusion. One limiting assumption of the QLT is that in the guiding center approximation transport of the GCR particles is not perturbed by the IMF turbulence. This assumption is inaccurate especially for the highly anisotropic strong turbulent heliosphere. This classical approach has been improved by higher-order theories of GCR particles turbulent flow. Thus the theories considering nonlinear effects have been introduced \cite{bib:matheus95}-\cite{bib:shalchi09}.  A validity of the QLT for the GCR particles of the energy $>1 GeV$ is confirmed by the weakly nonlinear theory (WNLT) \cite{bib:schalchi04a}, nonlinear parallel diffusion theory (NLPA) \cite{bib:Qin07} and in papers \cite{bib:droge03,bib:shalchi04b,bib:shalchi09} (see e.g. \cite{bib:WA10}).

However, for selecting  correct set of modulation parameters used in theoretical modelling (especially diffusion coefficients, etc.), one criterion remains the most important, if it is possible, to estimate them from the experimental data, supported by appropriate theory. 

Observations of GCR intensity and anisotropy by neutron monitors (NMs) and IMF fluctuations can be successfully used for establishment of various parameters characterizing modulation of GCR by the solar wind. In this paper we present the temporal evaluation of the parallel diffusion coefficient of GCR particles for rigidities to which NMs respond. Parallel diffusion coefficient $K_{\parallel}$, equivalent to parallel mean free path (MFP) $\lambda_{\parallel}=\frac{3K_{\parallel}}{v}$, determined by physical properties of interplanetary medium, is a very important parameter to study the transport of energetic particles in the heliosphere, especially for a solar event (SEP) connected with the space weather prediction \cite{bib:Qin09,bib:Qin11}.

Using data of the product $\lambda_{\parallel} \nabla_{r} n$ of the parallel MFP $\lambda_{\parallel}$  and radial gradient  $\nabla_{r} n$ of GCR calculated based on the GCR  anisotropy data (Ahluwalia et al., this conference ICRC 2013, poster ID: 487 \cite{bib:Ahl13}), we estimate the temporal changes of the radial gradient $\nabla_{r} n$  of GCR at the Earth's orbit. As a final point, determination of the parallel diffusion coefficient  $K_{\parallel}$ (equivalent to parallel MFP $\lambda_{\parallel}$) of GCR particles  according to Quenby \cite{bib:Quenby} and Hedgecock \cite{bib:hedg75} formulas will be performed for minimum conditions of solar activity.
\section{Convection-diffusion approximation}
It has been shown that $\sim 75-80\%$ of the 11-year variation of the GCR intensity can be interpreted based on the diffusion$-$convection model of GCR propagation \cite{bib:Dorman01,bib:Alania08}. So, the long-term variations of GCR intensity can be described by the Parker transport equation, invoking the isotropic convection-diffusion approximation \cite{bib:parker63}. In scope of this approximation one can  calculate  changes of the parallel diffusion coefficient $K_{\parallel}$  as follows:
\begin{eqnarray}
I=I_{0}exp(-\int_{r_{0}} ^{r_{E}} \frac{Vdr}{K_{\parallel}})\nonumber \\
dI=\frac{I_{0}-I}{I_{0}}\nonumber \\
dI\approx \int _{r_{0}} ^{r_{E}}\frac{Vdr}{K_{\parallel}}\nonumber \\
K_{\parallel}\propto\frac{Vr}{dI}
\end{eqnarray}
where  $dI$ is variation of the GCR intensity, $V$ solar wind velocity and  $r$ radial distance.
 \begin{figure}[t]
  \centering
  \includegraphics[width=0.45\textwidth]{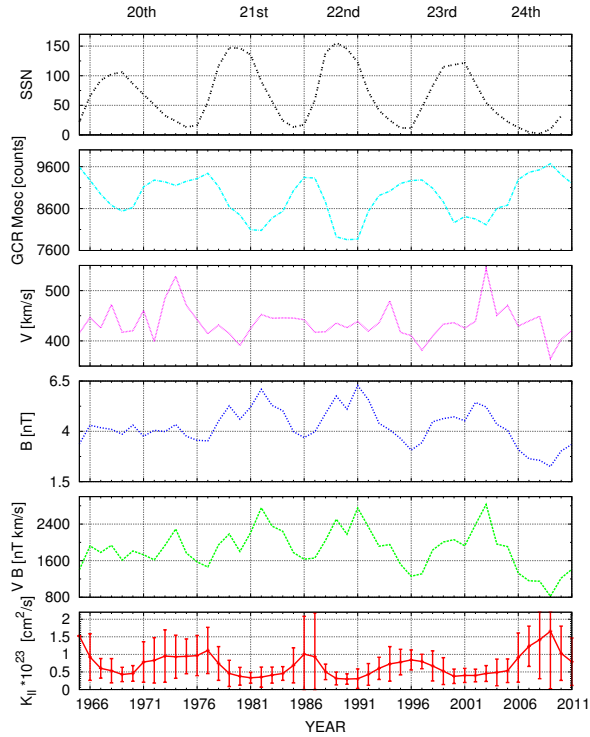}
  \caption{Temporal changes of the annual SSN, GCR intensity for Moscow NM, solar wind velocity $V$, magnitude $B$ of the IMF, product $V B$ and estimated parallel diffusion coefficient $K_{\parallel}$  for 1965-2011.}
  \label{simp_fig1}
 \end{figure}
 
Figure \ref{simp_fig1} shows a plot of the temporal changes of the annual sunspot numbers (SSN), GCR intensity for Moscow NM, the solar wind velocity $V$, the magnitude $B$ of IMF, product $V B$ and the estimated parallel diffusion coefficient $K_{\parallel}$  according to expression (1) for the period of 1965-2011. One notes that the parallel diffusion coefficient $K_{\parallel}$  exhibits  $\sim11$-year variation, but a stronger solar polarity dependence ($\sim22$-year variation); a significant increase is observed in the minimum epochs of solar activity, especially in the $A<0$ magnetic polarity period. An anomalous increase of  $K_{\parallel}$ for the recent solar minimum $23/24$ is clearly seen, as well.
\section{Parallel mean free path}
It has been suggested that parallel mean free path (MFP) $\lambda_{\parallel}$  can be expressed in terms of interplanetary magnetic field $B$ as follows \cite{bib:zank98}-\cite{bib:bazilevskaya13}:
 \begin{figure}[t]
  \centering
  \includegraphics[width=0.45\textwidth]{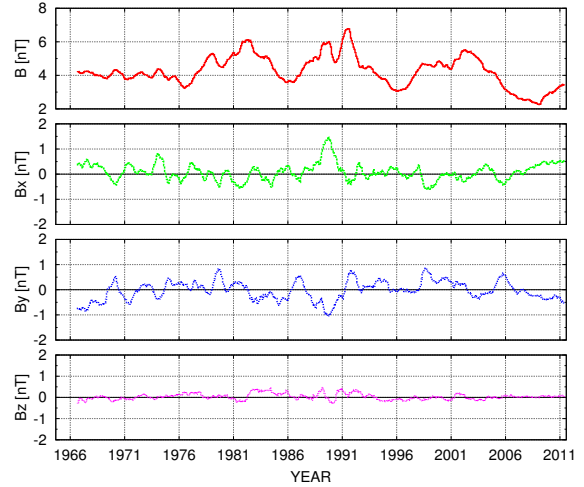}
  \caption{Temporal changes of the 13-month smoothed magnitude  $B$ and $B_{x}$, $B_{y}$ and $B_{z}$ components of the IMF for 1965-2011.}
  \label{simp_fig2}
 \end{figure}
 \begin{eqnarray}
\lambda_{\parallel}\propto \frac{B^{\frac{5}{3}}}{\delta B^{2}}
\end{eqnarray}
where $\delta B$  is the standard deviation. In our calculations the formula  (2) is replaced by an equivalent expression:
\begin{eqnarray}
\lambda_{\parallel}\propto \frac{B}{\delta B}
\end{eqnarray}
We calculate parallel MFP  $\lambda_{\parallel}$ from the $27-$day running averages of the observed radial $B_{x}$, azimuthal $B_{y}$, and latitudinal $B_{z}$ components of IMF according to the formulas:
\begin{eqnarray}
\lambda_{x}\propto \frac{B_{x}}{\delta B_{x}}, \lambda_{y}\propto \frac{B_{y}}{\delta B_{y}}, \lambda_{z}\propto \frac{B_{z}}{\delta B_{z}},
\end{eqnarray}
\begin{figure}[t]
  \centering
  \includegraphics[width=0.45\textwidth]{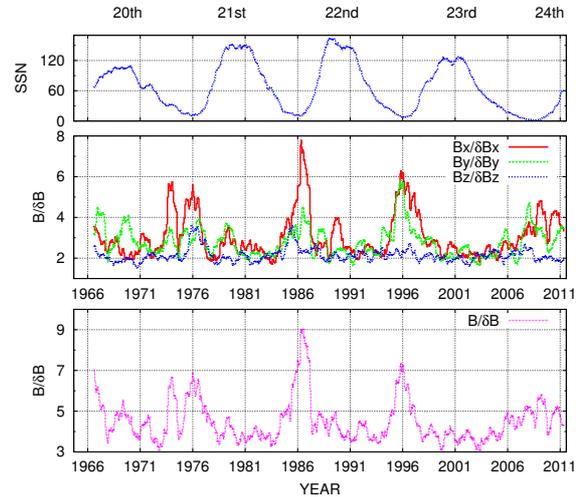}
  \caption{Temporal changes of the 13-month smoothed SSN and parallel MFP $\lambda_{\parallel}$  calculated according to expressions (3) and (4) for 1965-2011.}
  \label{simp_fig3}
 \end{figure}
 Data sets of the IMF magnitude $B$ and $B_{x}$, $B_{y}$, $B_{z}$ componets used in this calculation are presented in figure \ref{simp_fig2}. The results of MFP calculations for 13-month smoothed data for the corresponding components and magnitude of the IMF according to expressions (3) and (4) are presented in figure \ref{simp_fig3}. One notes that the MFP oscillates with a period of $\sim 11$ year solar activity cycle with a significant increase in the minimum periods of solar activity. Also, MFP is strongly polarity dependent in accord with the drift theory with a considerable enhancement especially in the minimum epoch of solar activity in the $A<0$ magnetic polarity period.

\section{Radial gradient of GCR}
\begin{figure}[t]
  \centering
  \includegraphics[width=0.45\textwidth]{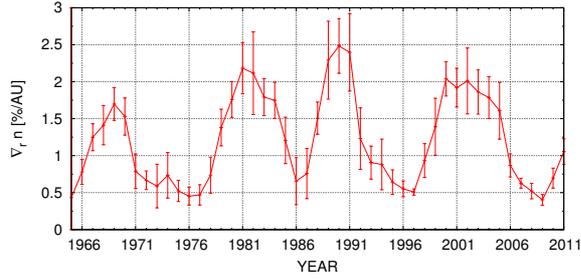}
  \caption{Time variation of the annual GCR radial gradient $\nabla_{r} n$  for 1965-2011 with errors bars.}
  \label{simp_fig4}
 \end{figure}

On the basis of the long term changes of the product $\lambda_{\parallel} \nabla_{r} n$  calculated based on the GCR  anisotropy data (Ahluwalia et al., this conference ICRC 2013, poster ID: 487 \cite{bib:Ahl13}) and parallel diffusion coefficient $K_{\parallel}$ of GCR found above (figure \ref{simp_fig1}), we estimate also the radial gradient $\nabla_{r} n$  of GCR at the Earth's orbit. The results of our calculations are presented in figure \ref{simp_fig4}. One notes that the radial gradient $\nabla_{r} n$  of GCR oscillates with a period $\sim 11$year solar activity cycle with a weaker solar polarity dependence, being in agreement with the previous calculation reported by Chen and Bieber \cite{bib:Chen93} and with PIONEER/VOYAGER observations \cite{bib:fuji97}.

\section{Transport coefficient for minimum conditions of solar activity}
Transport coefficients (e. g. diffusion coefficient of GCR) may be derived from a precise knowledge of the regular interplanetary magnetic field values and its fluctuations (turbulence) \cite{bib:Quenby}. The derivation of the parallel diffusion coefficient given by Jokipii \cite{bib:jokipii66,bib:jokipii67} and Hasselmann and Wibberenz \cite{bib:hasselman68} is best illustrated in a simple way by following the Kennel and Petschek \cite{bib:kennel66} formulation given in the context of magnetospheric particle scattering.

In order to accurately obtain the parallel mean free path (diffusion coefficient) of GCR particles in the heliosphere, a method of power spectrum density of the interplanetary magnetic field turbulence has been used \cite{bib:hedg75}. In this paper we compare the values of parallel MFP obtained by means of formulation of  Hedgecock \cite{bib:hedg75} and Quenby \cite{bib:Quenby}. We consider frequency range $10^{-6}-10^{-5} Hz$, responding for modulation of the GCR particles detected by NMs.

The appropriate parallel MFP can be expressed in terms of the power spectrum density of the interplanetary magnetic field fluctuations according to Hedgecock \cite{bib:hedg75} and Quenby \cite{bib:Quenby}, respectively:
\begin{eqnarray}
\lambda_{\parallel}\propto \frac{2\nu(\nu+2)cR^{2}}{9V \cdot P(f)}
\end{eqnarray}
\begin{eqnarray}
\lambda_{\parallel}\propto \frac{V B^{2}}{4 \pi \cdot P(f)} \frac{1}{f^{2}}
\end{eqnarray}
Where $V$ is the solar wind velocity, $B$ magnitude of IMF, $R$ - magnetic rigidity of GCR particles to which NM respond (in this case $R=15GV$), $f$ is the resonant frequency of GCR scattering, $P(f)$   is the power spectrum density at the resonant frequency $f$ with the spectral index $\nu$ for the frequency range $10^{-6}-10^{-5} Hz$.
 \begin{figure}[t]
  \centering
  \includegraphics[width=0.45\textwidth]{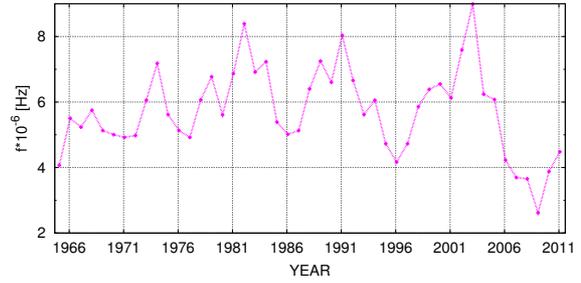}
  \caption{The time variation of the resonant frequency  $f$ for the 1965-2011 period.}
  \label{simp_fig5}
 \end{figure}
The time variation of the resonant frequency $f=\frac{V}{2 \pi}\frac{300 B}{R}$   for the period  1975-2011 is shown in figure \ref{simp_fig5}. Figure \ref{simp_fig5} shows clear $11-$year variation in accordance with the solar activity cycle.
 \begin{figure}[t]
  \centering
  \includegraphics[width=0.45\textwidth]{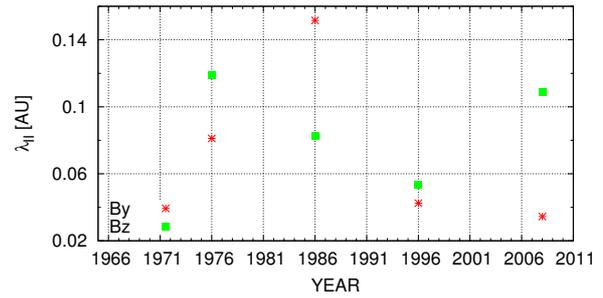}
  \caption{Values of the parallel mean free path  $\lambda_{\parallel}$  calculated based on expression (5) according to Hedgecock   \cite{bib:hedg75} for the consecutive minimum epochs of solar activity with different signs of global magnetic polarity for the period of 1975-2011. Each point corresponds   to the average value for three years around each solar minimum period.}
  \label{simp_fig6}
 \end{figure}

 \begin{figure}[t]
  \centering
  \includegraphics[width=0.45\textwidth]{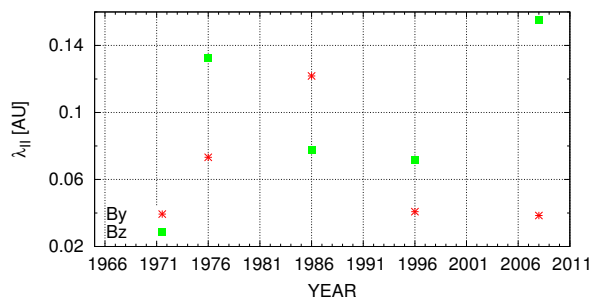}
  \caption{Values of the parallel mean free path  $\lambda_{\parallel}$  calculated based on expression (6) according to Quenby  \cite{bib:Quenby}  for the consecutive minimum epochs of solar activity with different signs of global magnetic polarity for 1975-2011. Each point corresponds   to the average value for three years around each solar minimum period.}
  \label{simp_fig7}
 \end{figure}
Figures \ref{simp_fig6} and \ref{simp_fig7} present the values of the parallel mean free path  $\lambda_{\parallel}$  calculated based on expression (5) and (6) according to Hedgecock  \cite{bib:hedg75} and Quenby \cite{bib:Quenby}, respectively, for the consecutive minimum epochs of solar activity with different signs of global magnetic polarity for the period of 1975-2011. The parallel mean free path  $\lambda_{\parallel}$ is calculated based on the transverse components $B_{y}$ and $B_{z}$  of the IMF. Each point corresponds   to the average value for three years around each solar minimum period (e.g., 1986 corresponds to the time interval 1985-1987).
Calculations for parallel MFP according to both formulas (5) and (6) are in good agreement with each other. One can see that parallel MFP calculated based on the $B_{y}$ transverse component of the IMF is strong polarity dependent with large increase in $A<0$ for 1985-1987. Unfortunately this statement for $B_{y}$ component is not satisfied in the last minimum 2007-2009 ($A<0$). On the other hand in the last minimum period with record level of the GCR intensity ever measured by NMs, calculations for $B_{z}$ transverse component show an increase in the changes of the parallel MFP.

\section{Conclusions}

\begin{enumerate}
\item The parallel diffusion coefficient  $K_{\parallel}$, obtained based on the isotropic convection-diffusion GCR modulation model,  generally displays  $\sim 11$-year variation, but with strong  polarity dependence ($\sim 22$ years). A significant increase is observed in the minimum epochs of solar activity, especially in the $A<0$ magnetic polarity period. An anomaly increase of $K_{\parallel}$  in recent solar minimum $23/24$ is clearly seen, as well.
\item We calculate parallel mean free path $\lambda_{\parallel}$ of GCR based on the experimental data of the IMF. Its value is polarity dependent in accord with drift theory and oscillate with a period of $\sim 11$-years solar activity cycle.
\item On the basis of the long term changes of the GCR anisotropy we show the $\sim 11$-year variation of the radial gradient $\nabla_{r} n$ of GCR being in good agreement with the PIONEER/VOYAGER observations.
\item Parallel mean free path $\lambda_{\parallel}$ calculated based on the $B_{y}$ transverse component of the IMF is strong polarity dependent with large increase in $A<0$ for 1985-1987. In the last minimum epoch of solar activity with record level of the GCR intensity ever measured by NMs, calculations for $B_{z}$ transverse component of the IMF show an increase in the changes of the parallel mean free path.
\end{enumerate}

\vspace*{0.5cm}
\footnotesize{{\bf Acknowledgment:}{ Author greatly benefited from discussions with Prof. M.V. Alania and Prof. H.S. Ahluwalia and thanks for assistance in evaluating this paper. Author thanks the investigators of neutron monitors stations and OMNI data base for
possibility to use their data.}}

\end{document}